\documentclass[twocolumn,showpacs,preprintnumbers,amsmath,amssymb]{revtex4}
\usepackage{graphicx}
\usepackage{dcolumn}

\usepackage[english]{babel}
\usepackage{amsmath}
\usepackage{amssymb}
\usepackage{tabularx}
\usepackage{multirow}
\usepackage[all]{xy}
\usepackage{appendix}
\usepackage{xspace}
\usepackage{color}

\begin{document}
\title{Harmonic generation and filamentation: when secondary radiations have primary consequences}

\author{P. B\'ejot$^{1}$} \email{pierre.bejot@u-bourgogne.fr}
\author{G. Karras$^{1}$}
\author{F. Billard$^{1}$}
\author{E. Hertz$^{1}$}
\author{B. Lavorel$^{1}$}
\author{E. Cormier$^{2}$}
\author{O. Faucher$^{1}$}

\affiliation{$^{1}$ Laboratoire Interdisciplinaire CARNOT de Bourgogne, UMR 6303 CNRS-Universit\'e de Bourgogne, BP 47870, 21078 Dijon, France}
\affiliation{$^{2}$ Centre Lasers Intenses et Applications, Universit\'e de Bordeaux-CNRS-CEA, UMR 5107, 351 Cours de la Lib\'eration F-33405 Talence, France}


\newlength{\textlarg}
\newcommand{\strike}[1]{%
  \settowidth{\textlarg}{#1}
  #1\hspace{-\textlarg}\rule[0.5ex]{\textlarg}{0.5pt}}


\begin{abstract}
In this Letter, it is experimentally and theoretically shown that  
 weak 
odd harmonics generated during the propagation of an infrared ultrashort ultra-intense pulse unexpectedly modify  the nonlinear properties of the medium and lead to a strong modification of the propagation dynamics. This result is in contrast with all current state-of-the-art propagation model predictions, in which secondary radiations, such as third harmonic, are expected to have a negligible action upon the fundamental pulse propagation. By analysing full three-dimensional \textit{ab initio} quantum calculations describing the microscopic atomic optical response, we have identified a fundamental mechanism resulting from interferences between a direct ionization channel and a channel involving one single ultraviolet photon. This mechanism is responsible for wide refractive index modifications in relation with significant variation of the ionization rate. This work paves the way to the full physical understanding of the filamentation mechanism and could lead to unexplored phenomena, such as coherent control of the filamentation by harmonic seeding. 
\end{abstract}

\maketitle

Since its first experimental observation in the mid-1990s \cite{Braun95}, laser filamentation, i.e., the nonlinear propagation of ultrashort intense laser pulses, has attracted a lot of interest in recent years because of its physical interest, as well as its important applications including few-cycle optical pulse generation, terahertz generation, supercontinuum generation, and remote sensing \cite{ChinReport,BergeReport,MysyReport,KasparianW08}. The main feature of filamentation is its ability to sustain very high intensities (around 50\,TW/cm$^2$) over very long distances in contrast with predictions of linear propagation theory. When exposed to such laser field intensities, atoms and molecules exhibit highly nonlinear dynamics leading to the observation of phenomena such as multiphoton and tunnel ionization, as well as harmonic generation. As far as the third-harmonic generation (THG) is concerned, it is now well known that this nonlinear process occurs during the filamentation process,  where about 1\,$\%$ conversion efficiency has been reported \cite{BergeTHG,AkozbekTHG}. Such a radiation, emitted in the ultraviolet for Ti:sapphire femtosecond lasers, is generally considered as having a negligible action on the propagation dynamics of filaments. This is because the framework describing the filamentation, based on ionization rate initially derived for a monochromatic wave, predicts that such secondary ultraviolet radiations have no impact on atom-field nonlinear dynamics as long as their intensities remain at the few percent level. The generated third harmonic therefore remains only a byproduct of the interaction that can be described as a first-order scattering process \cite{MoloScattering} and having negligible feedback on it. A few pioneer works \cite{BergeTHG,Bache2012} have nevertheless underlined the fact that third harmonic, accompanying the filamentation process, can saturate the nonlinear refractive index, leading to an effective high-order Kerr effect. This purely macroscopic effect, relying on phase matching, is then expected to slightly decrease the intensity within the filament core and increase the filament length. However, in all these studies, the underlying hypothesis is that THG does not modify the microscopic atomic response at the fundamental frequency. In parallel, a strong attention has been paid to the ionization of atoms and molecules in the strong field regime. In particular, a lot of studies, either theoretical or experimental, were devoted to two-color ionization, in which a laser and its harmonics participate to the ionization process. Already in the sixties, Popov et al. tackled analytically the problem of two-color ionization \cite{Popov67}. Specifically, it was predicted that third harmonic, when synchronized  with the fundamental laser, is responsible for a very strong ionization enhancement, which was confirmed by experiments in the multiphoton \cite{Charalambidis94} and tunneling regimes \cite{Watanabe94}. However, up to now, neither experimental nor theoretical study has addressed the influence of THG (and more generally harmonic generation) on the nonlinear optical properties of a medium in a microscopic point of view. In this Letter, we show that THG actively participates to the propagation dynamics of intense femtosecond laser by modifying the atomic response (and consequently, the refractive index) at the microscopic level, in contrast with all predictions made so far by state-of-the-art nonlinear propagation models. The strong variations of ionization induced by the UV field and its subsequent impact on the optical properties of the medium are experimentally investigated. This effect is then confirmed by \textit{ab initio} time-dependent 3D quantum calculations describing the microscopic response of atoms when submitted to a coherent two-color field. It is also theoretically predicted that higher-order harmonics impact even more the filamentation dynamics. Finally, using numerical nonlinear propagation simulations, it is shown that THG is sufficiently efficient at the beam focus of a moderate power laser to strongly modify its propagation dynamics. This conclusion completely modifies the filamentation model in which secondary radiations are considered as weak perturbations upon the propagation dynamics. This could potentially lead to unexplored phenomena, such as the coherent control of the filamentation of IR beams by UV/VUV seed beams.

An atom interacting  with an intense ultrashort laser can simultaneously absorb a large number of photons, leading to excitation or ionization where a fraction of the bound electronic wavepacket is promoted in the continuum. This excited atom exhibits different optical properties as compared to the initial atom, leading to the modification of the propagation of the laser pulse. In our experiment, the modification of the optical properties of argon induced by an intense ultrashort 790\,nm laser is assessed by the pump-probe defocusing technique described in \cite{RenardDefocusing,LoriotDefoc}. Details about the implementation of this technique can be found in the supplementary materials. One has to emphasize that the defocusing signal is proportional to $\Delta n^2$, i.e., the square of the peak to valley change of refractive index experienced by the probe beam. This was confirmed by 3D+1 numerical calculations simulating the experimental defocusing setup. With the probe beam following the pump, a long lived signal is observed. Since neither rotational nor vibrational effect exist for a mono-atomic gas, the post-pulse signal is the signature of a modification of the electronic structure of the atom induced by the pump. Namely, this signal is mainly due to electrons promoted into continuum states of the atom (ionization). The cross-defocusing signal recorded at large positive delay is thus proportional to the square of the amount of electrons promoted into the continuum. It then allows a direct experimental measurement of the ionization yield.

\begin{figure}[htbp!]
  \begin{center}
    \includegraphics[keepaspectratio, width=8cm]{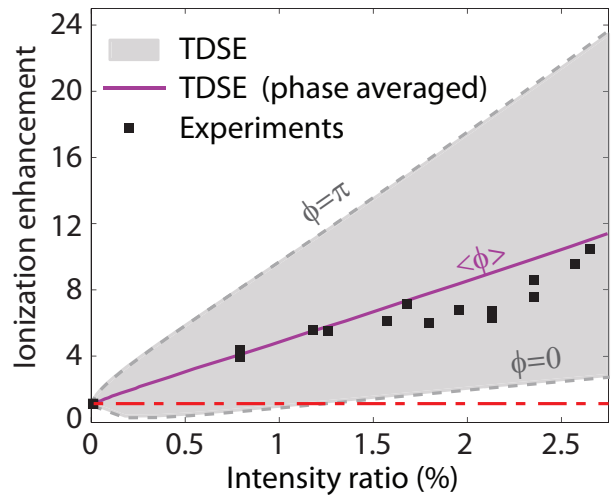}
  \end{center}
  \caption{(Color online) Experimental (black dots) and theoretical (magenta) ionization yield enhancement as a function of the ratio between UV and IR intensities for a 52 TW/cm$^2$ IR pulse. The theoretical curve is the average over all possible phases (represented as the gray area). The red dot-dashed line represents the ionization yield without the UV pulse.}
  \label{ExpTHG}
\end{figure}

Propagation models accounting for  both fundamental and third-harmonic radiations have been considered by several authors \cite{BergeTHG,AkozbekTHG,MysyTHG,CouaironTHG,MoloScattering}. The ionization resulting from the combination of a strong IR pulse and a weak THG pulse has been so far evaluated by adding the contributions of both pulses separately. As a consequence, because the UV intensity remains at the percent level when induced by the THG mechanism, the resulting ionization is almost the one induced by the IR beam alone. In order to test this hypothesis, a weak UV beam, generated by a non-collinear optical parametric amplifier and with central wavelength corresponding to the 
third harmonic, was spatially and temporally overlapped with the 
IR beam. The obtained ionization yield was then compared with the one induced by the single IR 
  beam. The intensity of the UV beam was set to be at the percent level of the IR beam so that no ionization was recorded when the latter was blocked. When the UV pulse was temporally shifted away from the IR pulse, no difference could be observed compared to the IR-pump measurements. Conversely, when the UV and the IR pulses were synchronized, a spectacular increase of the ionization yield was recorded. For a UV intensity of 1\,\%, the ionization increased by a factor of 4. This result, in complete disagreement with state-of-the-art propagation models, indicates that interferences between ionization paths involving IR and UV photons occur. In order to assess the physical mechanism underlying this coherent effect, the UV intensity was changed keeping the IR intensity constant. As shown in Fig. \ref{ExpTHG}, the enhancement of the ionization yield increases almost linearly with the ratio between the  UV and the  IR intensities, suggesting that the interference path inducing such an increase involves a single UV photon.
\begin{figure}[htbp!]
  \begin{center}
    \includegraphics[keepaspectratio, width=8cm]{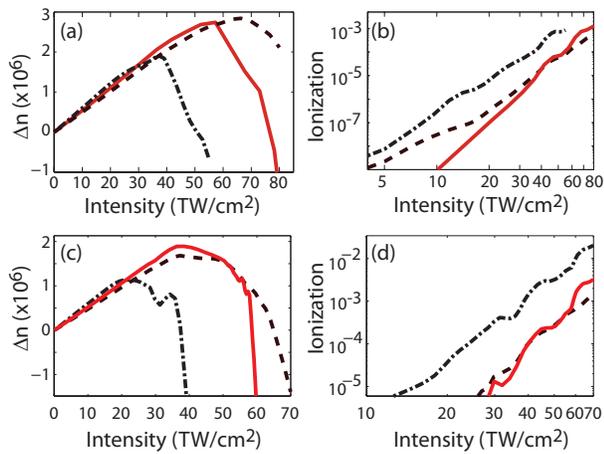}
  \end{center}
  \caption{(Color online) Left: Nonlinear refractive index $\Delta n$ \textit{vs} pump intensity without (red solid line) and with 1\,\% of third harmonic  for different relative phases : $\phi=0$ (black dashed) and $\phi=\pi$ (black dash-dot). Right: Ionization yield after the interaction as a function of the peak intensity. The pulse duration are 23\,fs (a,b) and 93\,fs (c,d), respectively.}
  \label{ResultsTHGTheo1}
\end{figure}
In order to confirm these results, time-dependent 3D quantum calculations describing the atom-strong field interaction at the microscopic level were performed. The calculations, based on the time-dependent Schr\"{o}dinger equation (TDSE) describing the interacting atom, were used to evaluate both the population promoted into the continuum after the interaction and the nonlinear refractive index as done in \cite{BejotPRL3}. Calculations were performed in argon under the single-active electron approximation. Further details about the calculation methods are provided in the supplementary materials. Ionization yield and nonlinear refractive index induced by a single IR 
 beam were evaluated as a function of intensity for two distinct pulse durations: a short 23\,fs and a long 93\,fs pulse (full width at half maximum). As shown in Fig. \ref{ResultsTHGTheo1}, for the IR electric field alone, the nonlinear refractive index increases linearly with the peak intensity in the low-field regime as expected for a purely cubic Kerr effect. For the short (long) pulse, it saturates at 50\,TW/cm$^2$ (35\,TW/cm$^2$), and finally becomes negative around 80\,TW/cm$^2$ (60\,TW/cm$^2$). This result is in line with what has been calculated recently in other atomic systems \cite{BejotPRL3,Nurhuda1,Nurhuda2,Kano,Volkova,TelekiWK2010}. Now when THG is added to the IR field with a relative intensity of 1\,$\%$ only, the nonlinear refractive index experienced by the latter drastically changes, as shown in Fig. \ref{ResultsTHGTheo1}. While almost no change is noticeable in the weak-field regime, the intensity at which the nonlinear refractive index becomes negative can either increase or decrease, depending on the relative phase between the two pulses. This behavior is correlated with the ionization yield change induced by the addition of the UV field as shown in Fig. \ref{ResultsTHGTheo1}(b). For intensities below 40\,TW/cm$^2$, adding 1\,\% of UV radiation strongly increases the ionization yield for any phase. The situation becomes more complex in the strong field regime. While a significant increase is still noticeable when the UV and IR pulses are out of phase, destructive interferences take place when the two pulses are in phase, leading to a drop of the ionization yield. One has to emphasize here, that the experimental setup was not sensitive to the relative phase between the IR  and the UV pulses but only to the phase-averaged signal. As shown in Fig. \ref{ExpTHG}, an excellent agreement is found between the phase-averaged ionization yield enhancement predicted by full TDSE calculations and the experimental results.

\begin{figure}[tb!]
  \begin{center}
    \includegraphics[keepaspectratio, width=8cm]{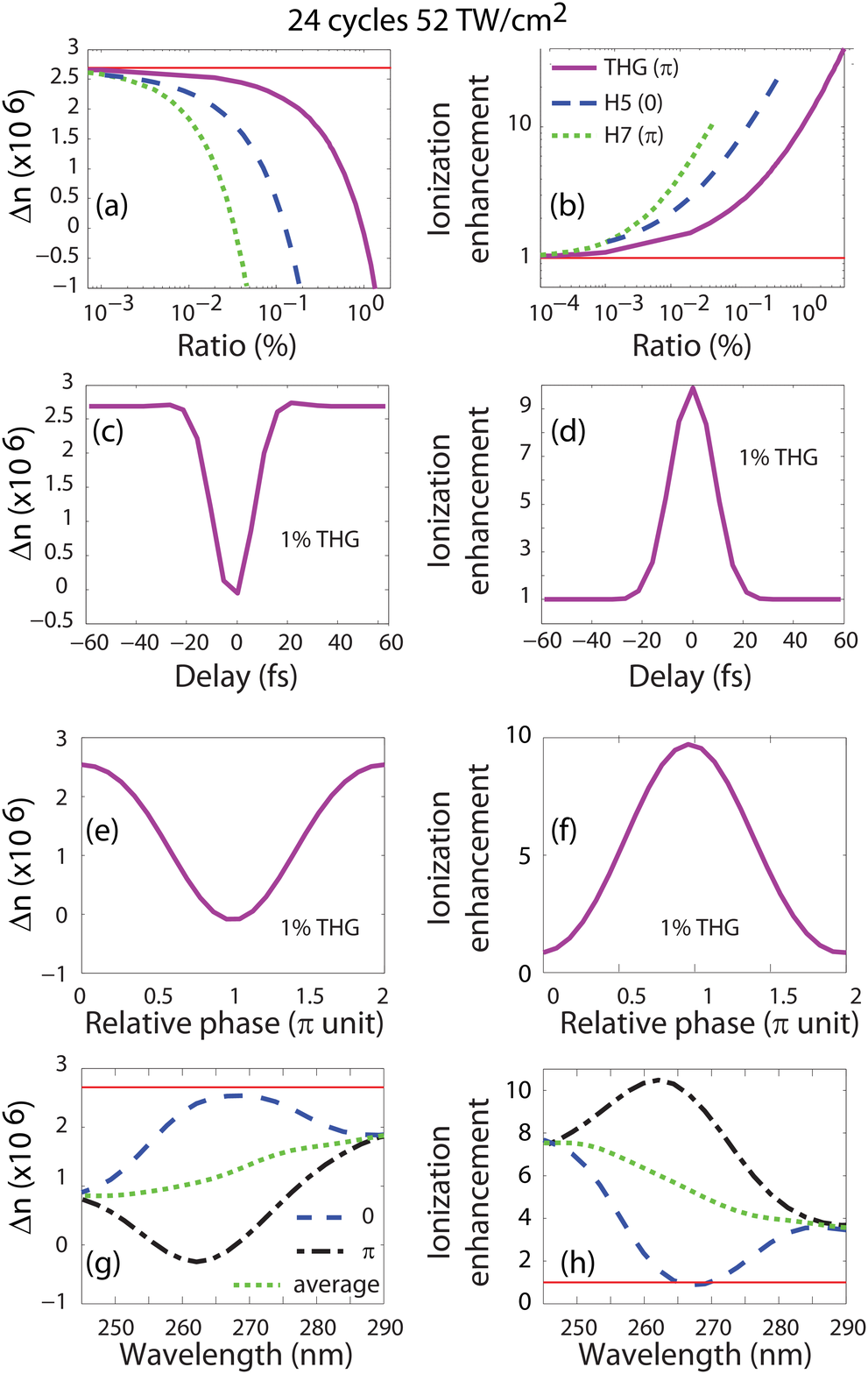}
  \end{center}
  \caption{(Color online) Nonlinear refractive index $\Delta n$ (a) and ionization enhancement (b) \textit{vs} intensity ratio between the third (THG), fifth (H5), and seventh (H7) harmonics and the fundamental for a 24 cycles, 52\,TW/cm$^2$ pulse. Nonlinear refractive index (c) and ionization enhancement (d) \textit{vs} delay between a 52\,TW/cm$^2$ pulse and 1\,\% of THG. In panels (a-d), the relative phase has been set to $\pi$, excepted for the H5 case in panels (a,b) where the relative phase is 0. (e,f): The same \textit{vs} 
  relative phase between 1\,\% of THG and IR pulses. (g,h): The same \textit{vs} 
  central wavelength of the 1\,\% UV pulse and a relative phase 0 (blue dashed lines) and $\pi$ (black dash-dot lines). The green dotted lines correspond to the  phase averaged values. The red solid lines are the level obtained for the single IR pump case.}
  \label{ResultsTHGTheo2}
\end{figure}

The influence of the intensity ratio, relative delay, and relative phase, which are the three parameters that will most likely modify the interference mechanism taking place when the IR and UV pulses are mixed, have been numerically studied. The results, summarized in Fig. \ref{ResultsTHGTheo2} for the short pulse, have been obtained for a 52\,TW/cm$^2$ IR beam, i.e., at the intensity where the nonlinear refractive index induced by the single IR pulse is maximal. Four important points are confirmed by the calculations. First, the ionization yield depends linearly on the intensity of the UV beam, as already shown in Fig. \ref{ExpTHG}. The process consequently involves a single UV photon. Second, the effect is optimal when the UV and IR beams temporally overlap, as shown in Figs. \ref{ResultsTHGTheo2}(c,d). Third, we see from Figs. \ref{ResultsTHGTheo2}(e,f) that the maximal ionization enhancement is observed for a phase $\phi_{\textrm{max}}$ of $\pi$ and a minimum for a phase $\phi_{\textrm{min}}$ of $0$. This is because the total IR+UV peak field is slightly higher for $\phi=\phi_{\textrm{max}}$ than for $\phi=\phi_{\textrm{min}}$. As the ionization process is highly nonlinear, a slight increase of 10 \% in the field amplitude results in a variation as large as an order of magnitude in the ionization. Note that the exact values of $\phi_{\textrm{max}}$ and $\phi_{\textrm{min}}$ depend on the definition used for the electric field. For instance, while $\phi_{\textrm{max}}=\pi$ and $\phi_{\textrm{min}}=0$ for sinus carrier electric fields, which corresponds to the electric field definition used in the present calculations (see the supplementary materials), $\phi_{\textrm{max}}=0$ and $\phi_{\textrm{min}}=\pi$ for cosinus carrier electric field. Finally, the impact of the central wavelength of the weak UV pulse has been investigated [see Figs. \ref{ResultsTHGTheo2}(g,h)]. While an effect on both nonlinear refractive index and ionization yield is noticeable for any wavelength, the phase dependence vanishes as soon as the UV spectrum is not anymore within the third-harmonic bandwidth. In this case, the relative phase between the two incommensurable fields varies during the pulse, leading to a time-averaged signal independent of the phase. The results of Fig. \ref{ResultsTHGTheo2} clearly indicate that the enhancement of the ionization yield originates from interference between UV and IR fields. The influence of higher-order harmonics on both ionization yield and refractive index has also been numerically investigated. As it is the case for the third harmonic, the ionization yield enhancement depends linearly on the fifth and seventh harmonic intensity. Nevertheless, as shown in Figs. \ref{ResultsTHGTheo2}(a-b), the influence of these harmonics is even more drastic on both ionization yield and nonlinear refractive index. Indeed, while about 1\,\% of UV is necessary to drop the nonlinear refractive index down to zero for a 52\,TW/cm$^2$ IR intensity, the same results is achieved with only 0.1\,\% of the fifth harmonic or 0.04\,\% of the seventh harmonic.
Note that, contrary to the third and seventh harmonics, the optimal phase for enhancing the ionization yield with the fifth harmonic is 0 as the total peak field amplitude is now obtained for this phase value. More generally, one can show that the optimal relative phase for a weak $2k+1$ order harmonic is $k\pi$ for sinus carrier electric fields as used in the present calculations.

\begin{figure}[htbp!]
  \begin{center}
    \includegraphics[keepaspectratio, width=8cm]{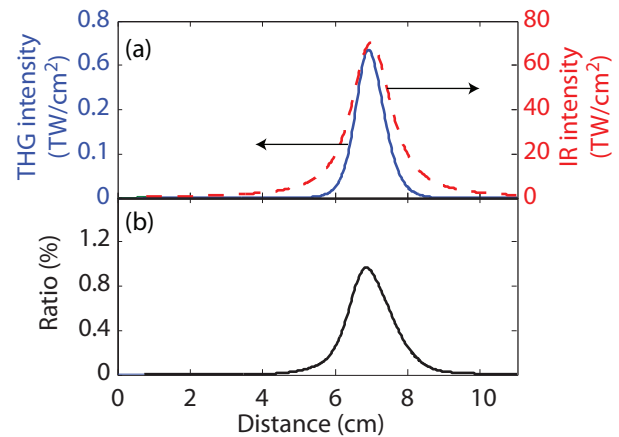}
  \end{center}
  \caption{(Color online) On-axis intensity (a) of the 
  IR (red dashed line) and THG (blue solid line) pulses.  Ratio between the THG and pump intensities (b) as a function of the propagation distance.}
  \label{Propagation}
\end{figure}

The experimental results presented in this work have been obtained  with an external UV beam. In order to estimate if self-induced THG can influence the propagation of an IR beam, the THG efficiency was calculated along the nonlinear propagation of an intense IR beam with the help of the unidirectional propagation equation \cite{UPPE_Moloney}. Details on the nonlinear propagation equation can be found in the supplementary materials. Figure \ref{Propagation}(a) shows the on-axis intensity of a 40\,$\mu$J, 23\,fs laser fundamental pulse, calculated  along its propagation in 1\,bar of argon, together with the generated third harmonic. The numerical simulation starts 7\,cm before the position of the 40\,$\mu$m waist. As shown in Fig.\ref{Propagation}(b), the ratio between THG and IR intensities increases until it reaches the focal point, where it starts to decrease due to the Gouy phase shift experienced by the IR beam, as mentioned in \cite{MysyTHG}. The fact that the ratio reaches the percent level at moderate power (\textit{i.e.} around 0.15 critical power \cite{Fibich}), clearly indicates that the interference effect can have a deep impact on the propagation dynamics of high power laser filament. In particular, it has been shown \cite{AkozbekTHG,MysyReport} that the relative phase between a filament and its subsequent third harmonic is locked to $\pi$, so that the effect of ionization enhancement is expected to be maximal. 
One must emphasize that this effect depends nonlinearly with pressure. In particular, since both nonlinear refractive index, responsible for the THG mechanism and phase matching conditions depend on pressure, it is anticipated that the THG-IR interaction  will be significantly decreased for experiments performed in low gas pressure cells or molecular jets, where nonlinear propagation have marginal effects.

In conclusion, we have shown that, contrary to the prevailing belief shared by  the filamentation community, a realistically weak UV beam co-propagating with an intense IR beam can actively modify the nonlinear properties of the medium experienced by the latter, leading to a strong modification of its propagation dynamics. This feedback mechanism is mainly attributed to ionization quantum-pathways mixing one single ultraviolet-photon with infrared photons and can lead to either an increase or a decrease of the ionization rate, as compared to the one induced by the infrared pulse alone. It is also expected that higher-order harmonics impact even more the filamentation dynamics. By evaluating the THG efficiency during the propagation of a moderate power IR pulse, it is shown that the exhibited UV-IR interference effect could deeply impact the propagation dynamics of a filament. This result is in contrast with all current state-of-the-art propagation model predictions, in which secondary radiations, such as third harmonic, are expected to have a negligible feedback upon the fundamental-pulse propagation. In particular, this work suggests that odd harmonics have to be taken into account in the modeling of filamentation, especially when evaluating the ionization rate. This work paves also the way to the full physical understanding of the filamentation mechanism and could potentially lead to unexplored phenomena, such as the coherent control of ionization and, in turn, of the nonlinear propagation of infrared beams by UV/VUV seed beams.

\bibliographystyle{srt}

\acknowledgments
This work was supported by the Conseil R\'egional de Bourgogne (PARI program), the CNRS, and the Labex ACTION program (contract ANR-11-LABX-01-01). E.C. and P.B. thank O. Peyrusse and the CRI-CCUB for CPU loan on their respective multiprocessor servers. The authors gratefully acknowledge J. Kasparian for very fruitful discussion.\vfill

\end{document}


\title{Harmonics generation and filamentation: supplementary materials}
\maketitle
\section{Time-dependent Schrodinger equation}
Within the dipole and single active electron approximations, the 3D time-dependent Schrodinger equation (TDSE) describing the evolution of the electron wavefunction $\ket{\psi}$ in the presence of an electric field $\textbf{E}(t)$ reads:
\begin{equation}
i\frac{d\ket{\psi}}{dt}=(H_0+H_{\textrm{int}})\ket{\psi},
\end{equation}
where $H_0=\mathbf{\nabla}^2/2+V_{\textrm{eff}}$ is the atom Hamiltonian, $H_{\textrm{int}}=\textbf{A}(t)\cdot\boldsymbol{\pi}$, where $\textbf{A}(t)$ is the vector potential such that $\textbf{E}(t)=-\partial \textbf{A}/\partial t$ and $\boldsymbol{\pi}=-i\mathbf{\nabla}$) is the interaction term expressed in the velocity gauge. The effective potential $V_{\textrm{eff}}$ of argon is the one originally proposed in \cite{MullerPotential}. It accurately fits the argon eigen-energies and -wavefunctions. The time-dependent wavefunction $\ket{\psi}$ is expanded on a finite basis of B-splines allowing memory efficient fast numerical calculations with a very large basis set \cite{Bachau2001}:
\begin{equation}
  \psi(\textbf{r},t) = \sum_{l=0}^{l_\textrm{max}}\sum_{i=1}^{n_\textrm{max}} c_{i}^{l}(t)
  \frac{B_{i}^k(r)}{r}Y_{l}^{0}(\theta,\phi),
\label{eq:TDSE_expansion}
\end{equation}
where $B_i^k$ and $Y_l^m$ are B-spline functions and spherical harmonics, respectively. The basis parameters ($l_\textrm{max}$, $n_\textrm{max}$, $k$ and the spatial box size) and the propagation parameters are chosen to ensure convergence. The atom is initially in the ground state (3p) and the electric field $E$ is linearly polarized along the $z$ axis and is expressed as $\textrm{E}(t)=E_0\cos^2(t/\sigma_\textrm{t})\sin(\omega_0t)$ for $|t|<\pi N/\omega_0$, where $\omega_0$ is the central frequency of the laser, $\sigma_\textrm{t}=2N/\omega_0$, and $N$ the total number of optical cycles within the pulse. The simulations are performed for a laser wavelength of 800\,nm and pulse durations (FWHM of intensity) of 23\,fs (93\,fs) corresponding to  $N$=24 ($N$=96) cycles. The third harmonic electric field is expressed as $\textrm{E}_{\textrm{TH}}(t)=\sqrt{R}E_0\cos^6[(t-\tau)/\sigma_\textrm{t}]\sin[3\omega_0(t-\tau)+\phi]$, where $\phi$ ($\tau$) is the relative phase (delay) between fundamental and third harmonic electric fields and $R$ is the relative intensity of the third harmonic with respect to the fundamental.

The microscopic polarization $p_z(t)$ along the $z$ axis and the medium polarizability $\alpha$ at the driving frequency $\omega_0$ are defined in atomic units as:
\begin{eqnarray}
p_z(t)&=&\bra{\psi(t)}r\cos\theta\ket{\psi(t)}\\ 
\alpha(\omega_0)&=&\frac{p_z(\omega_0)}{E(\omega_0)}, 
\end{eqnarray}
where $p_z(\omega)$ [$E(\omega)$] is the Fourier transform of $p_z(t)$ [$E(t)$] and $\theta$ is the angle between the dipole moment and its projection along the electric field polarization. Under the isolated atom assumption, the macroscopic polarization $P_z$ and susceptibility $\chi$ are then calculated as $P_z(t)=\mathcal{N}p_z(t)$ and $\chi=\frac{\mathcal{N}}{\epsilon_0}\alpha$, respectively, where $\mathcal{N}$ is the number density. The refractive index $n$ is then calculated as $n=\sqrt{1+\chi}$ and the nonlinear refractive index $\Delta n$ as:
\begin{equation}
\Delta n=n(I_0)-\lim_{I_0 \ \mapsto 0 } \limits n(I_0),
\end{equation}
where $I_0$ is the peak intensity of the pulse.

\section{Propagation and third harmonic generation}

Assuming a cylindrical symmetry around the propagation axis, the equation driving the propagation of the linearly polarized electric field envelope $\mathcal{E}$ embedding both fundamental and THG pulses reads in the reciprocal space \cite{UPPE_Moloney}:
\begin{eqnarray}
\partial_z\widetilde{\mathcal{E}}&=&i\left(k_z-k_1\omega\right)\widetilde{\mathcal{E}}+\frac{1}{k_z}\left(\frac{i\omega^2}{c^2}\widetilde{P}_{\textrm{NL}}-\frac{\omega}{2\epsilon_0c^2}\widetilde{J}\right)\nonumber\\
& & -\widetilde{L}_{\textrm{losses}},
\end{eqnarray}
where $\widetilde{P}_\textrm{NL}$ is the nonlinear polarization including both Kerr effect and third harmonic generation, $\widetilde{J}$ is the plasma-induced current, $\widetilde{L}$ is the nonlinear losses expressed in the reciprocal space, and
\begin{eqnarray}
k(\omega)&=&n(\omega)\omega/c,\nonumber\\
k_z&=&\sqrt{k^2(\omega)-k^2_r},\nonumber\\
P_{\textrm{NL}}&=&n_{2}\left(|\mathcal{E}|^{2}+\frac{1}{3}\mathcal{E}^2e^{2iw_0t}\right)\mathcal{E},\nonumber\\
\widetilde{J}&=&\frac{e^2}{m_\textrm{e}}\frac{\nu_\textrm{e}+i\omega}{\nu_\textrm{e}^2+\omega^2}\widetilde{\rho\mathcal{E}},\nonumber\\
\partial_t\rho&=&W\left(|\mathcal{E}|^{2}\right)(\rho_\textrm{at}-\rho),\nonumber\\
L_{\textrm{losses}}&=&\frac{\rho_\textrm{at}W\left(|\mathcal{E}|^{2}\right)}{2|\mathcal{E}|^{2}},\nonumber
\end{eqnarray}
where $\omega$ is the angular frequency, $n$ is the frequency dependent refractive index, $c$ is the light velocity, $k(\omega)$ is the frequency dependent wavenumber, $k_1=\partial k/\partial \omega|\omega_0$ corresponds to the inverse of the group velocity, $k_r$ is the conjugate variable of $r$ in the reciprocal space, $m_\textrm{e}$ and $e$ are the electron mass and charge respectively, $n_2$ is the nonlinear refractive index, $\rho$ ($\rho_{\textrm{at}}$) is the electron (atom) density, $W$ is the ionization rate derived by Popov et al. \cite{PPT}, and $\nu_\textrm{e}$ is the effective collision frequency.

\section{Experimental setup}

\begin{figure}[htbp!]
  \begin{center}
    \includegraphics[keepaspectratio, width=8cm]{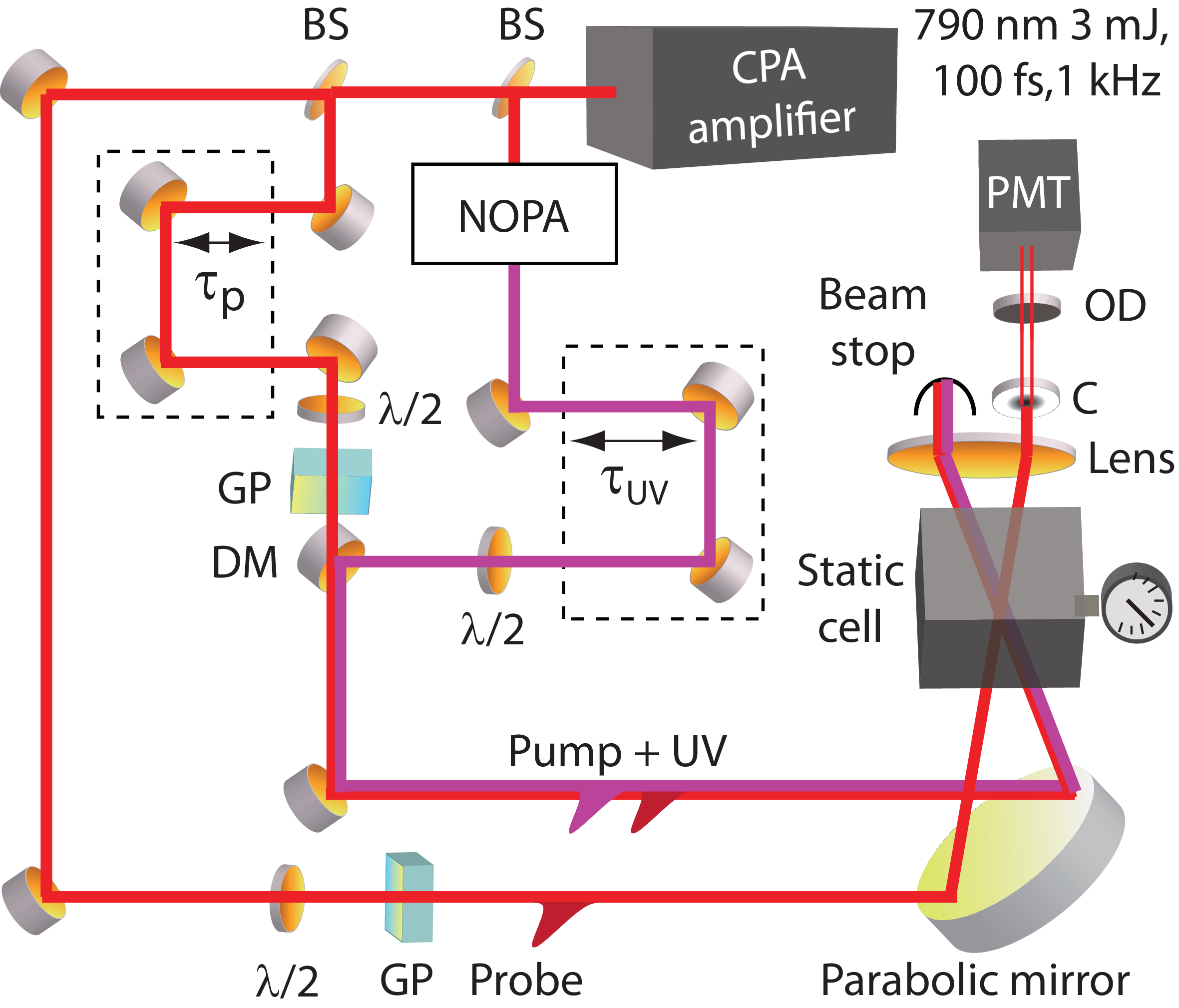}
  \end{center}
  \caption{(Color online) Pump-probe setup for laser-induced cross-defocusing measurements. BS: Beam Splitter, GP: Glan Polarizer, OD: Neutral Optical Density, PMT: Photo Multiplier Tube, DM: Dichroic mirror, C: Coronagraph. The noncollinear optical parametric amplifier (NOPA) is used for generating the UV beam.}
  \label{setupTHG}
\end{figure}

The experimental setup used for the presented measurements is depicted in Fig. \ref{setupTHG}. The optical source is a 1\,kHz amplified femtosecond laser delivering horizontally polarized, 3\,mJ, 100\,fs pulses at 790\,nm.  The refractive index change induced by the pump beam in a static gas cell filled with argon at 0.4\,bar is measured by the pump-probe cross-defocusing technique described in \cite{RenardDefocusing,LoriotDefoc}. All the beams are focused with a $f$=15\,cm off-axis parabolic mirror to avoid both longitudinal and lateral chromatism. A third beam with central wavelength corresponding to the pump third harmonic is generated by means of a non collinear parametric amplifier (NOPA) and spatially superimposed with the pump beam with the help of a dichroic mirror. The temporal synchronization of the UV beam with respect to the pump is accomplished with the help of a delay line. The polarization of all beams is set vertical with half-wave plates. The pump (probe) energy is adjusted at 50\,$\mu$J (0.5\,$\mu$J) with the help of zero-order half-wave plates and glan polarizers while the one of the UV beam is adjusted by controlling the energy of the input IR beam of the NOPA. The IR and UV peak intensities are estimated from the measurement of their respective waists and durations. After the coronagraph, the remaining part of the probe beam is redirected to a photomultiplier tube and then integrated with a boxcar. The recorded signal is averaged over 200 laser shots for increasing the signal-to-noise ratio. Since the defocusing signal induced in the UV+IR scheme is about 50 times higher than the one induced during the single pulse experiment, a calibrated neutral optical filter is added before the photomultiplier in this case in order to keep the detection dynamics constant and to avoid a saturation of the photomultiplier. The defocusing signal is proportional to $\Delta n^2$, i.e., the square of the peak to valley change of refractive index experienced by the probe beam. When the pump pulse precedes the arrival of the probe, the cross-defocusing signal is proportional to the square of refractive index change resulting from the ionization mechanism and accordingly proportional to the square of the amount of ionized electrons generated by the pump. It then allows a direct experimental measurement of the latter. Finally, since third harmonic and pump pulses are generated from the same laser source, their relative phase should be constant from shot to shot. Nevertheless, their respective optical paths are distinct over about 4 m up to the recombining dichroic mirror. As a consequence, because of random fluctuations of their respective optical paths (due to air motion, mechanical vibrations...) and because no active stabilization is used, their relative phase varies from shot to shot which leads to an averaging over all possible relative phases of the cross-defocusing signal when the latter is integrated over 200 shots.

\bibliographystyle{srt}